\documentclass{article}
\usepackage{amsmath}
\usepackage{graphicx}
\usepackage{color}
\usepackage{multirow}
\usepackage{comment}
\usepackage{url}

\title{Analysis of the Impact of an Execution Algorithm with an Order Book Imbalance Strategy on a Financial Market Using an Agent-based Simulation}

\providecommand{\keywords}[1]
{
  \small	
  \textbf{\textit{Keywords---}} #1
}

\author{%
Shuto Endo$^{1}$, Takanobu Mizuta$^{2}$, Isao Yagi$^{3}$  \\
        \small $^{1}$Information and Computer Sciences, Graduate School of Engineering, Kogakuin University \\
        \small $^{2}$SPARX Asset Management Co., Ltd. \\
\small $^{3}$Department of Information Science, Faculty of Informatics, Kogakuin University
}

\begin{document}

\maketitle
\begin{abstract} 
One measure of financial markets is order book imbalance (OBI), defined as the number of buy orders minus the number of sell orders in the order book around the best quote. Thus, OBI is an indicator of an imbalance in supply and demand, which can potentially impact the price as price movements in the financial market. There is known to be a positive correlation between OBI and returns, and some investors attempt to exploit this fact to improve their investment performance. 
In financial markets, an investor sometimes wants to place a large order. However, when an investor places such a large order all at one time, some other investors learn that someone wants to trade a large amount and themselves trade earlier than they would have, or the market price becomes volatile and losses are incurred.
Therefore, execution algorithms are used, by which a machine automatically divides orders into smaller lots to avoid distorting the market price. Execution algorithms could be expected to achieve improved performance when OBI is taken into account, but no such findings have been reported so far. One reason for the lack of research findings is that it is virtually impossible to measure the impact of OBI alone on the performance of execution algorithms, since there are many external factors that can affect financial markets. Multi-agent simulation is one way to solve problems that are difficult to analyse using empirical research and conventional methods. In a multi-agent simulation, individual actors in the world are regarded as agents and the behavioural rules and interactions of these agents constitute a model. In this context, a financial market constructed using multi-agent simulation is called an artificial market. In the present study, we modelled an execution algorithm considering OBI, investigated how the model is affected by the market under several market patterns using artificial markets, and analysed the mechanisms. The results showed that in stable markets, the performance of the execution algorithm with the OBI strategy varied with the number of orders placed. In contrast, in markets with unstable prices, the performance of the execution algorithm with the OBI strategy was higher than that of the conventional execution algorithm. Even in markets with manipulation by spoofing, the performance of the execution algorithm with the OBI strategy was not significantly worse than that of the conventional execution algorithm, demonstrating that the model is not easily affected by spoofing.
\end{abstract}


\begingroup
\renewcommand\thefootnote{}
\footnotetext{This manuscript is an English translation of the authors' original article published in the \textit{Journal of the Japanese Society for Artificial Intelligence} (Vol. 39, No. 4, 2024). The original Japanese version is available at: \url{https://www.jstage.jst.go.jp/article/tjsai/39/4/39_39-4_FIN23-I/_article/-char/en}.}
\addtocounter{footnote}{-1}
\endgroup

% 本文
\section{Introduction}
%\section{まえがき}
Participants in financial markets, including but not limited
to asset-management firms, often wish to place large orders
involving high volumes of securities.
However, any single high-value order
is a signal to the market that there exist investors 
wishing to make high-value trades; this may
provoke other participants to buy or sell earlier than they would have,
distorting market prices and potentially causing losses.
The goal of avoiding such price distortions motivates the
use of execution algorithms, by which
machines conduct trades by automatically subdividing 
large orders into smaller portions to be placed
sequentially~\cite{CJ16,Ma20,SS19}.
Maechler~\cite{Ma20} noted that, in recent years, execution algorithms
have evolved from elementary approaches that simply subdivided large orders
over fixed intervals to complex strategies that account for 
shifting market conditions when placing orders.\par

In stock markets, the difference between the numbers of buy and sell orders
in the vicinity of the best quote is known as the order book imbalance (OBI).
OBI is known to be positively correlated with returns; namely, 
positive returns are obtained when buy orders exceed sell orders
near the best quote, whereas negative returns are obtained when sell
orders exceed buy orders~\cite{CHW09,CJP15,CKS14,St17,LPS13}.
Empirical studies have shown that strategies that account for the
properties of OBI (OBI-aware strategies) can increase profits~\cite{CDJ18,Ru15}.
For example, Cao et al.~\cite{CHW09}
analysed data from the Australian Securities Exchange and
found that OBI was related to future returns.
Similarly, Cartea et al.~\cite{CDJ18}
used data on Nasdaq-traded brand-name stocks to
test the performance of an OBI-aware strategy and found that the strategy significantly
increased profits, demonstrating the usefulness of OBI for 
predicting price fluctuations.
These findings suggest that the performance of execution algorithms
may be improved by adopting OBI-aware strategies; however, a review of the securities research 
literature found no previous empirical studies investigating this question.\par

Characterizing the performance of a given strategy in
actual markets is complicated by various external factors,
independent of the strategy in question and differing from 
market to another, whose influence must be excluded 
for the purposes of securities analysis.
One technique capable of resolving questions
that are difficult to address via conventional
methods of empirical studies is
multi-agent simulation, which models the
behaviour of individual investors, as agents, in the world subject to
rules governing their actions and interactions.
Financial markets constructed using such multi-agent simulations
are referred to as artificial markets~\cite{AHLPT97,CCD12,CIP09,Ye13}.
The goal of using artificial markets is not to achieve faithful quantitative reproduction
of phenomena occurring in actual financial markets, but rather to
observe sequences of processes that result from changes in a given
environment to see what types of consequences ensue and what sorts of mechanisms drive them.
For this reason, studies to characterize the performance of strategies
tend to use models that are as concise as possible while being 
designed to facilitate the analysis of simulation results.
Artificial markets have been used to study how markets are influenced by
market regulations~\cite{MS99,ZPZLX16},
and there have even been studies of machine learning
algorithms for agents in artificial markets~\cite{HMIS20}.
However, although OBI-aware strategies using artificial markets
have been investigated~\cite{YHM23},
there have been no studies to date of
the performance of execution algorithms using OBI-aware strategies.\par

Therefore, in the present study, we constructed artificial markets
reflecting four distinct categories of market environment---a stable
market, a crash market in which a flash crash occurs,
a surge market in which prices rise sharply, and
a spoofing market in which spoofing manipulation occurs---and 
investigated for each 
how market conditions affect the performance of execution algorithms,
based on OBI-aware strategies.
Section 2 describes our artificial-market models and the various
agents they contain, and Section 3 outlines our experimental procedure.
Section 4 presents our simulation results and discusses their ramifications.
Finally, Section 5 summarizes our conclusions. This study constitutes an
update, based on additional experiments, of previous work 
by Endo et al.~\cite{EMY23b}.

%\section{人工市場モデル}
\section{Artificial market model}
In this study, we constructed a model based on the artificial markets of Endo et al.~\cite{EMY23a} and Yagi et al.~\cite{YHM23}, and newly incorporated an execution algorithm based on the algorithm agent of Hoshino et al.~\cite{HMY21}.
The number of normal agents (NAs) was set to 990, and the number of algorithm agents was set to 10.
NA $j$ (NA $j$) 1 to 990 submit orders for one share at a time in sequence, where after the order by NA 990 is submitted, NA 1 submits the next order. Conventional algorithm agent (AA) $k$ decides whether to place an order every $l$ periods, where $k$ is incremented every $l$ periods and after $k=10$, the count returns to $k=1$. The time step $t$ is incremented by one each time an NA places an order or an AA decides whether to place an order. Even if an order is submitted but no trade is executed, time $t$ is still advanced by one step. All orders submitted by an agent are given a cancel time $t_c$. When time $t_c$ has elapsed since the order was placed, the order is cancelled and removed from the order book.

The pricing mechanism in the model is a continuous double auction. This means that if there are buy (sell) order prices in the order book that are higher (lower) than the sell (buy) order price of the agent, then an agent's order is immediately matched to the highest buy order (lowest sell order) in the order book. On the other hand, if there are no orders in the order book, then the order does not match any other order and remains in the order book. We refer to the former type of order placed by an agent as a market order and the latter type as a limit order. Among orders at the same price, priority is given to those placed earlier. Thus, the rules of price priority and time priority are applied in the continuous double auction used in our experiment.
%
%本研究では遠藤ら\cite{EMY23a}とYagiら \cite{YHM23}の人工市場を基に，新たに星野ら\cite{HMY21}のアルゴリズムエージェントを参考にした執行アルゴリズムを加えてモデルを構築した．
%\par
%一般投資家エージェントは990体，アルゴリズムエージェントは10体とする．
%一般投資家エージェント$j$は$j=1$から順に注文を出し，$j=990$までが注文を終えた後，再び$j=1$から順に注文を行う．
%アルゴリズムエージェント$k$は，$l$期ごとに順に注文を出すか判断し，$k=10$まで到達すると再び$k=1$に戻る．
%時刻$t$は一般投資家エージェント1体が注文を出す，または執行アルゴリズム1体が注文を出すか判断するたびに1だけ増える．注文をしただけで取引が成立しない場合も時刻$t$は1ステップ進む．また，各エージェントが発注したすべての注文には有効期限$t_c$が設けられている．発注された時刻$t$から有効期限$t_c$が経過すると注文が破棄され，注文板上から削除される．
%\par
%このモデルの価格決定メカニズムは，買い手と売り手が価格を提示し，両者の価格提示が合致するとその価格で取引が成立するザラバ方式（連続ダブルオークション方式）とした．このザラバ方式では投資家が出した注文と注文板の状況を比較し，投資家の買い（売り）注文価格より低い（高い）最良売り（買い）注文が注文板に存在すれば成行注文となり取引が成立する．存在しなければ投資家の注文は指値注文となり注文板に残る．また，同じ価格の注文は先に出された注文から取引が行われるように注文板を形成している．よって，本実験のザラバ方式では価格優先・時間優先のルールが適用されている．

\subsection{Normal agents}
%\subsection{一般投資家エージェント}

\newcommand{\RE}{${re}^t_j$} % 予想リターン
\newcommand{\WIJ}{${w_i}^t_j$} % 重み　一般化
\newcommand{\WIMAX}{${w_i}_{\mathrm{max}}$} % 1,2項目目の重みの最大値
\newcommand{\WONE}{${w_1}^t_j$} % 1項目目の重み
\newcommand{\WTWO}{${w_2}^t_j$} % 2項目目の重み
\newcommand{\WTHREE}{$u_j$} % 3項目目の重み
\newcommand{\UMAX}{$u_{\mathrm{max}}$} % 3項目目の重みの最大値

NAs are assumed to be general investors in the real world and are designed to be able to replicate the characteristics of the real markets. An NA submits a buy or sell order with the order price which is determined by combining the following three trading strategies: the fundamental strategy, the technical strategy, and the noise strategy. The fundamental strategy refers to the fundamental price to make investment decisions. The technical strategy uses past price movements to make investment decisions. Finally, the noise strategy represents trial-and-error investment decisions. The weights of the fundamental and technical strategies are changed by learning as market conditions change.

We now explain the order process and the learning process of NAs. In the order process, NA $j$ decides the order prices as follows. The rate of the expected price of NA $j$ at
time $t$, i.e., the expected return \RE, is given by the following equation:

%一般投資家エージェントは一般的な投資戦略に基づいて取引を行う投資家を想定したエージェントであるが，現実の市場における一般的な注文状況を再現する役割も担っている．一般投資家エージェントはファンダメンタル価格を参照し投資判断を行うファンダメンタル戦略，過去の価格推移を利用して投資行動を行うテクニカル戦略，試行錯誤的な投資判断を表すノイズ戦略の3種の戦略を組み合わせて発注価格を決め，買いまたは売りの注文を出す．また，市場の状況に合わせて学習することでファンダメンタル戦略とテクニカル戦略の比重を適宜切り替えていく．本エージェントは遠藤ら \cite{EMY23a}とYagiら\cite{YHM23}のモデルに基づいて構築した．

%以下に一般投資家エージェントの注文プロセスと学習プロセスを記す．まずは注文プロセスから述べる．
%一般投資家エージェントは以下の手順に従い，買いと売りの判断を行う．
%一般投資家エージェント$j$が時刻$t$のときに予想する価格の変化率（予想リターン）\RE は式(\ref{exp_return})から求められる．

\begin{eqnarray} 
	{re}^t_j=\frac{{w_1}^t_j{r_1}^t_j+{w_2}^t_j{r_2}^t_j+u_j\epsilon_j^t}{{w_1}^t_j+{w_2}^t_j+u_j}.
	\label{exp_return}
\end{eqnarray}

In equation (\ref{exp_return}), \WONE\ is the fundamental strategy weight, and \WTWO\ is the technical strategy weight. Both strategy weights are decided according to the uniform distribution between $0$ and \WIMAX\ at the beginning of the simulation. Each weight is changed
using the learning process described later. \WTHREE\ is the noise strategy weight and is decided according to the uniform distribution between 0 and \UMAX\ at the beginning of the simulation. We model the strategy weights as random variables chosen independently.
The denominator of equation (\ref{exp_return}) normalizes the influence of these components.
%
%右辺の分子の\WONE はファンダメンタル戦略の成分の重み，\WTWO はテクニカル戦略の成分の重み，\WTHREE はノイズ戦略の成分の重みであり，これらの重みはすべて実数値をとる．\WONE と\WTWO はシミュレーション開始時にそれぞれ$0$から\WIMAX までの一様乱数で決める．これらの重みは学習プロセスにて互いに独立して変化する．
%\WTHREE はシミュレーション開始時にそれぞれ0から\UMAX までの一様乱数で決められ，シミュレーション中は変化しない．
%これら3つの重みからくる影響は式(\ref{exp_return})の右辺の分母にて正規化することで平準化している．\par

\newcommand{\RI}{${r_i}^t_j$} % リターン 一般化
\newcommand{\RONE}{${r_1}^t_j$} % 1項目目のリターン
\newcommand{\RONEF}{$\ln{\left({P_f}/{P^{t-1}}\right)}$} % 1項目目のリターンの式
\newcommand{\RTWO}{${r_2}^t_j$} % 2項目目のリターン
\newcommand{\RTWOF}{$\ln{\left({P^{t-1}}/{P^{t-1-\tau_j}}\right)}$} % 2項目目のリターンの式
\newcommand{\TAUJ}{$\tau_j$} % 3項目目のリターン
\newcommand{\TAUMAX}{$\tau_{\mathrm{max}}$} % 3項目目のリターンの最大値

Also in equation (\ref{exp_return}), \RI\ is the expected return of a strategy, indexed by $i$, of NA $j$ at time $t$. Among these, \RONE\ denotes the expected return of the fundamental strategy, calculated as $\ln$ ($P_f$/$P_{t-1}$), which means that NA $j$ expects a positive (negative) return when the fundamental price is higher (lower) than the previous market price, where $P_t$ is the market price at time $t$ and $P_f$ is the fundamental price, which is constant over time. The initial market price $P_0$ is set to $P_f$, and the market price is set to the most recent price when no trades have been made. \RTWO\ denotes the expected return of the technical strategy and is calculated as $\ln$ ($P_{t-1}$/$P_{t-1-\tau_j}$), which means that NA $j$ expects a positive (negative) return when the historical return is positive (negative). Note that \TAUJ\ is taken from the uniform distribution between $1$ and \TAUMAX\ at the start of the simulation for NA $j$. $\epsilon_j^t$ is set as a normally distributed random
error with a mean of $0$ and standard deviation of $\sigma_\epsilon$.

The expected price of NA $j$ at time $t$, ${P_e}^t_j$, is calculated as follows:

%\RI は時刻$t$における一般投資家エージェント$j$の$i$項目の予想リターンである．1項目の\RONE はファンダメンタル成分のリターンであり，\RONEF とする．これは，ファンダメンタル価格と1期前の市場価格を比較し，市場価格の方が低ければ正，高ければ負の予想リターンを意味する．
%$P_f$は時間で変化しない一定のファンダメンタル価格である．ファンダメンタル価格とは，株式を発行する企業自体がもっている実態価値に基づいた価格のことである．$P^t$は時刻$t$における市場価格（取引されなかった時刻では直近取引された価格であり，$t=0$では$P^t=P_f$とする）である．2項目の\RTWO はテクニカル成分の予想リターンであり，\RTWOF とする．これは，過去のリターンが正なら正，負なら負の予想リターンを意味している．
%\TAUJ は1から\TAUMAX までの一様乱数でエージェントごとに決める．$\epsilon_j^t$は時刻$t$におけるエージェント$j$のノイズ成分であり，平均0，標準偏差$\sigma_\epsilon$の正規分布乱数である．\par
%式(\ref{exp_return})で導いた予想リターンを基に予想価格${P_e}^t_j$を式(\ref{exp_price})で求める
%	\footnote{
%		本研究では対数リターンを使用している．そのため予想リターンは現在の価格の対数と予想価格の対数の差である．すなわち，${re}^t_j=\ln{{P_e}^t_j}-\ln{P^{t}}=\ln{{P_e}^t_j/P^{t}}$であり，これより式(\ref{exp_price})が導き出される．
%}．

\begin{eqnarray} 
	{P_e}^t_j=P^{t-1}\mathrm{exp}\left({re}^t_j\right)
	\label{exp_price}
\end{eqnarray}

The order price of NA $j$ at time $t$, ${P_o}^t_j$, is set as normally distributed with a mean of ${P_e}^t_j$ and standard deviation of ${P_\sigma}^t_j$. Note that ${P_\sigma}^t_j={P_e}^t_j\cdot est$ and $est = 0.003$. 
If ${P_o}^t_j$ is less than ${P_e}^t_j$, then NA $j$ submits a buy order whose price is ${P_o}^t_j$ for one share. If ${P_o}^t_j$ is greater than ${P_e}^t_j$, then NA $j$ submits a sell order whose price is ${P_o}^t_j$ for one share.

%注文価格${P_o}^t_j$は平均${P_e}^t_j$，標準偏差${P_\sigma}^t_j$の正規分布乱数で決める．ただし，${P_\sigma}^t_j={P_e}^t_j\cdot est$とする．
%$est$は標準偏差を決めるための係数であり，今回は$est=0.003$とした．
%さらに${P_o}^t_j$が呼値単位$\Delta P$より小さい値をもつ場合，呼値より小さい部分を売り注文では切り上げ，買い注文では切り捨てる．例えば，$\Delta P=1$のとき，${P_o}^t_j$が小数点以下を持つ場合は，売り注文では小数点以下は切り上げ，買い注文では切り捨てる．
%そして，${P_o}^t_j$ が ${P_e}^t_j$ より小さければ，リスク資産1単位の買い注文を出し，${P_o}^t_j$ が ${P_e}^t_j$ より大きければ，リスク資産1単位の売り注文を出す．\par

The learning process is performed before the order process at each time point. Comparing the sign of \RI\ ($i=1,2$) with that of the return of learning period ${r_l}^t=\ln{\left({P^{t-1}}/{P^{t-1-t_l}}\right)}$, if the two signs are the same, then ${w_i}^t_j$ is updated to ${w_i}^t_j+k_l\left|{r_l}^t\right|{q_j}^t\left({w_i}_{\mathrm{max}}-{w_i}^t_j\right)$, where $t_l$ is a constant that determines the past time of the price that the agent refer to, $k_l$ is a constant that specifies the degree of adjustment applied to the component weight \WIJ, and ${q_j}^t$ is set according to the uniform distribution between 0 and 1. If \RI\ and ${r_l}^t$ have opposite signs, then \WIJ\ is updated to ${w_i}^t_j-k_l\left|{r_l}^t\right|{q_j}^t{w_i}^t_j$.
These rules mean that the weights of strategies whose predicted direction of price change coincides with the actual direction of price change are raised, whereas the weights
of strategies that are out of line are lowered. Furthermore, \WIJ\ is reset according to the uniform distribution between 0 and \WIMAX\ with probability $\delta _l$. This feature makes an objective model of the search for a better strategy by trial and error to fundamentally reevaluate the previous strategy.

%一般投資家エージェントの学習はエージェントごとに注文の直前に行われ，各成分の予想リターン\RI （ただし，$i=1,2$）の符号の向きと学習期間のリターン${r_l}^t=\ln{\left({P^{t-1}}/{P^{t-1-t_l}}\right)}$の符号の向きとを比較し，成分の重み\WIJ（ただし，$i=1,2$）を変更していく．予想リターン\RI と学習期間のリターン${r_l}^t$が同符号なら成分の重み${w_i}^t_j$を${w_i}^t_j+k_l\left|{r_l}^t\right|{q_j}^t\left({w_i}_{\mathrm{max}}-{w_i}^t_j\right)$に，異符号なら${w_i}^t_j-k_l\left|{r_l}^t\right|{q_j}^t{w_i}^t_j$に書き換えを行う（$||$は絶対値）．$t_l$は過去のどの地点の価格を参照するかを定める定数，$k_l$は成分の重み\WIJ をどの程度増減させるかを決定する定数，$q_j^t$は時刻$t$，エージェント$j$に与えられる0から1までの一様乱数である．価格変化の方向の予測と現実の価格変化の方向が一致した戦略の重みを引き上げ，外れている戦略の重みを引き下げるようになっている．またこれとは別に，\WIJ を確率$\delta _l$で0から\WIMAX までの一様乱数にて再設定も行う．この再設定はこれまでの投資戦略を抜本的に見直すために，試行錯誤的により良い戦略を模索している姿を客観的にモデル化したものである．

\subsection{Algorithm agents}
Algorithm agents are assumed to be the execution algorithms operating in real markets, and are implemented as follows. We prepared two types of algorithm agents: conventional algorithm agents (AAs) and OBI-strategy algorithm agents (OAAs). An AA places a market buy order for one share, whereas an OAA compares the buy depth (the number of buy orders within a certain range below the best bid) with the sell depth (the number of sell orders within a certain range above the best ask) and if the buy depth is greater than the sell depth, then the OAA submits a market buy order for one share and otherwise does not place any order. In this study, we analyse the differences in impact by introducing these two types of execution algorithms into the artificial market. Similar results can be obtained in cases where the algorithm agents place only market sell orders.

To equalize the number of order submissions between the two types of algorithm agents, we first execute a simulation with the OAAs and record the number of orders submitted by OAAs. Then, we set the order interval for the AAs so that their number of orders matches that of the OAAs and run the simulation. Furthermore, the algorithm agents start submitting orders after time step 100,000.

%\subsection{アルゴリズムエージェント}
%アルゴリズムエージェントは執行アルゴリズムを想定したエージェントである．
%アルゴリズムエージェントは以下のようにモデル化した．\par
%アルゴリズムエージェントは従来型アルゴリズムエージェント(AA)とOBI戦略型アルゴリズムエージェント(OAA)を用意した．AAは注文数1の成行買い注文を出す．OAAは市場の買いDepth（最良買い気配値から一定のティック下までの間に存在する買い注文数）と売りDepth（最良売り気配値から一定のティック上までの間に存在する売り注文数）を比較し，買いDepthが大きれば注文数1の成行買い注文を出し，そうでないときは注文を出さない．本研究では用意した2種類の執行アルゴリズムをそれぞれ人工市場に参入させることで，影響の違いを分析する．アルゴリズムエージェントが成行売り注文のみを行う場合も趣旨に沿った結果が得られると考えられる．\par
%2種類のアルゴリズムエージェントで発注回数をそろえるため，まずOAAを参入させた状態でシミュレーションを実行し，OAAの発注回数と同等の発注回数となるような発注周期を設定したうえでAAを参入させシミュレーションを行う．また，アルゴリズムエージェントは時刻100,000以降から発注を行う．

\section{Simulation}
%\section{実験}
In this study, we built four market simulation environments: 
\begin{enumerate}
\item a market with stable price fluctuations (stable market),
\item a market in which a flash crash occurs (crash market),
\item a market in which prices rise sharply (surge market), and
\item a market in which spoofing manipulation occurs (spoof market).
\end{enumerate}
In each of these environments, we observe the trading costs when the two types of algorithm agents (AA and OAA) participate. Note that the trading cost is calculated based on the trades of all algorithm agents (ten in total), not on the trades of a single algorithm agent.

The market environments are reproduced as follows. 

The stable market represents a market in which prices remain stable without erroneous orders. The crash market (surge market) is reproduced by having NAs, during 30,000 periods from time $100,000$ to $130,000$, submit sell (buy) orders at a price of $1$ ($100,000$) with a probability of $20\%$. Finally, the spoof market is created by repeating cycles in which spoofing is introduced into the stable market. Specifically, starting from time $100,000$, a certain number of spoofing orders are placed near the best bid in the order book. Periods where spoofing orders are added alternate with periods where no spoofing orders at all are added, with the switch occurring every $10,000$ periods. In this study, the number of spoofing orders is set to 1,000. In all market environments, the fundamental price is kept constant throughout the simulation.

The initial parameter values are shown in Table \ref{param}. $t_e$ denotes the time at which the simulation ends. Each simulation was run 25 times under each condition, and the results were used for the subsequent discussion.

%本研究では4つの市場環境を想定する．１つ目は価格変動が安定した市場(安定市場)で，２つ目はフラッシュクラッシュが起こった市場（急落市場），3つ目は市場価格が急上昇した市場（急騰市場），4つ目は見せ玉による相場操作が行われた市場（見せ玉市場）である．
%以上の市場環境において提案した2種類のアルゴリズムエージェント（AAとOAA）がそれぞれ参入した際の取引コストを観察する．なお，取引コストはそれぞれのアルゴリズムエージェント1体の取引ではなく全体（10体）の取引によって算出されたものである．\par
%市場環境は以下のようにモデル化することで再現する．\par
%安定市場は，誤発注などがない市場価格が安定した市場である．
%次に急落（急騰）市場は，一般投資家エージェントに時刻$100,000$から$130,000$にかけての$30,000$期の間，20\%の確率で注文価格1(100,000)の売り（買い）注文を出させることで急落（急騰）を再現する．
%最後に見せ玉市場は，安定した市場に対して時刻$100,000$から$10,000$期ごとに見せ玉を最良買い気配値付近の注文数に上乗せしていく期間と見せ玉操作を行わない期間を繰り返す．今回は見せ玉数を1,000とした．いずれの市場も，ファンダメンタル価格はシミュレーション期間全体を通して一定である．\par
%
%各パラメータの基準値を\ref{param}に示した．$t_e$はシミュレーション終了時の時刻である．
%シミュレーションは各条件の下でそれぞれ$25$回ずつ試行し，その結果をもとに議論を行った．

\begin{table}[t]
	\centering\caption{Parameters.\label{param}}
	\begin{tabular}{cc}
	Parameter & Initial value \\ \hline
	${w_1}_{\mathrm{max}}$ & 1 \\
	${w_2}_{\mathrm{max}}$ & 10 \\
	${u}_{\mathrm{max}}$ & 1 \\
	${\tau}_{\mathrm{max}}$ & 10,000 \\
	${\sigma}_{\epsilon}$ & 0.06 \\
	$\Delta P$ & 1 \\
	${P}_{f}$ & 10,000 \\
	${t}_{l}$ & 10,000 \\
	${t}_{c}$ & 20,000 \\
	$t_e$ & 400,000 \\
	${k}_{l}$ & 4.0 \\
	$\delta _l$ & 0.01 \\ 
	
	\end{tabular}
\end{table}

\subsection{Validation of proposed model}
%\subsection{人工市場モデルの妥当性}

\begin{table}[t]
	\centering\caption{Stylized facts ($\theta _{pm}=0.300\%$)
	\label{Stylized facts}}
	\begin{tabular}{c|cc}
	Kurtosis & \multicolumn{2}{c}{3.207189} \\ \hline
	& lag 1 & 0.1738 \\
	& lag 2 & 0.1067 \\
	& lag 3 & 0.0811 \\
	& lag 4 & 0.1242 \\
	\multirow{-5}{*}{Autocorrelation coefficients for squared returns} & lag 5 & 0.0277 \\ 
	\end{tabular}
\end{table}

Real financial markets have particular statistical properties, as empirical studies have confirmed~\cite{Co01,Se11}. However, as noted by Sewell~\cite{Se11}, only two stylized facts are observed stably and have significant effects across different periods: fat tails and volatility clustering. Other stylized facts are unstable, and attempting to reproduce them by making the model more complex may hinder the analysis of mechanisms. Therefore, we verified only whether our model can reproduce these two stylized facts. Regarding fat tails, in a histogram plotted based on data of price changes, the distribution has a high kurtosis and a thicker tail than a normal distribution. In general, a fat tail occurs when kurtosis is positive. Volatility clustering, on the other hand, measures the autocorrelation of squared returns. When volatility clustering is present, a positive correlation is observed even with lagged autocorrelation. The parameters of this model are set to make the model simple and stable, based on the research by Endo et al.~\cite{EMY23a} and Yagi et al.~\cite{YHM23}, to facilitate the analysis of experimental results. 

Table \ref{Stylized facts} shows the stylized facts in a market where OAAs participate. The statics for stylized facts are averages over 25 simulation runs. Note that lags in Table \ref{Stylized facts} have different degrees of shift when calculating the autocorrelation of squared returns; specifically, lag $n$ represents the autocorrelation between times $t$ and $t-n$. Table \ref{Stylized facts} shows that both the autocorrelation coefficients for squared returns with several lags and kurtosis are positive, which means that all runs replicated volatility clustering and a fat tail. Thus, the model reproduces long-term statistical
characteristics observed in actual financial markets.

Next, we checked whether OBI is correlated with future returns in our model. In a single simulation run, if the value obtained by subtracting the number of times when the buy depth is smaller (greater) than the sell depth and the price rises (falls) in the next period from the number of times when the buy depth is greater (smaller) than the sell depth and the price rises (falls) in the next period is positive, it can be concluded that there is a positive correlation between OBI and future returns. The results were positive in the stable market for both the case with AA participation and that with OAA participation.

Therefore, it can be confirmed that the proposed model is valid, as it reproduces both stylized facts observed in actual financial markets and the above property of OBI.

%実際の金融市場には統計的性質（スタイライズド・ファクト）が現れることが多くの実証研究によって示されている\cite{Co01,Se11}．しかし，Sewell\cite{Se11}が述べているように，どのような時期にも安定的に有意に観測されるスタイライズド・ファクトはファット・テールとボラティリティ・クラスタリングの2つしかない．その他のスタイライズド・ファクトは不安定であり，それらを再現するためにモデルを複雑化させるとメカニズム分析を阻害する恐れがあることから，これら2つのスタイライズド・ファクトのみを再現できているか検証した．ファット・テールとは，価格の変化率のデータから度数分布(ヒストグラム)を作成した際，正規分布に比べ，分布の先端が尖り(尖度が大きい)，分布の裾が厚くなっている状態である．尖度が正のとき，ファット・テールが成立している．ボラティリティ・クラスタリングとは，リターンが大きく変化した後には，しばらく大きな変化が続き，小さな変化の後には小さな変化が続く傾向があることを示す．これは定量的にはリターンおよびリターンの2乗の自己相関にラグがある場合でも正値になることで示される．\par
%本モデルでは，遠藤ら \cite{EMY23a}とYagiら \cite{YHM23}のモデルに基づき，スタイライズド・ファクトを再現しつつ可能な限りモデルが安定的に動作するようパラメータを設定した．
%\ref{モデルの妥当性}は，一般投資家エージェントとOAAが参入したときのスタイライズド・ファクトであり，25回のシミュレーションの平均値である．\ref{モデルの妥当性}で用いられているlagはリターンの2乗の自己相関をとる際のずらす度合を示している．尖度とリターンの2乗の自己相関の両方が正値となっているためスタイライズド・ファクトが現れていることを示している．つまり，使用したモデルが実際の金融市場で観察される統計的性質を再現しているので，本テーマを検証するためのモデルとして妥当であると判断した．
%
%次に，構築した人工市場モデルにおいて，OBIと将来のリターンに正の相関があるか検証した．買いDepthが売りDepthより大きい（小さい）かつその次の期に価格が上昇（下落）した場合を＋1，買いDepthが売りDepthより小さい（大きい）かつその次の期に価格が上昇（下落）した場合を－1としてシミュレーション期間全体でカウントした．その値が正であればOBIと将来のリターンに正の相関があるといえる．結果は安定市場においてAA参入時，OAA参入時の両方の場合で正となった．よって，構築した人工市場においてOBIに将来のリターンと正の相関があることが示された．\par
%以上より，本モデルでは実際の金融市場で観察される統計的性質およびOBIの特性が再現できているため，本テーマを検証するためのモデルとして妥当であると判断した．
%

\subsection{Trading cost}
%\subsection{取引コスト}
To calculate the average buy price for algorithm agents, we utilize trading costs, which serve as an indicator of how much an agent's own orders influenced the market price. Following Hoshino et al.~\cite{HMY21}, we define the trading cost $TC$ as the degree to which algorithm agents buy at prices higher than the fundamental price $P_f$, expressed as the average deviation of the buy price from the fundamental price, as follows:
%アルゴリズムエージェントの平均購入価格を求めるために取引コストを利用する．取引コストは自分自身の注文がどれくらい市場価格に影響を与えたかを示す指標である．
%今回は星野ら\cite{HMY21}の案に基づいて，アルゴリズムエージェントがファンダメンタル価格$P_{f}$よりどれだけ高い価格で取引を行ったかを取引コストTCと定義した（式(\ref{mi})参照）．取引コストはファンダメンタル価格に対する購入価格の平均乖離度で表している．

\begin{eqnarray} 
TC=\frac{1}{n_{buy}}\sum_{l=1}^{n_{buy}}{p_{buy}^l-P_f},
	\label{mi}
\end{eqnarray}
where $n_{buy}$ is the total number of purchases made by algorithm agents during the simulation, and $p_{buy}^l$ is the price at the $l$-th purchase. Note that $TC$ is calculated by aggregating the transactions of all algorithm agents. It is known that in the absence of algorithm agents, the average transaction price is approximately equal to the fundamental price \cite{MMKIKY14}. Therefore, the smaller the value of $TC$, the lower the average buy price achieved by the algorithm agents, meaning that their performance is superior.
%ここで，$n_{buy}$はシミュレート全期間を通してアルゴリズムエージェントが資産を購入した回数，$p_{buy}^l$は$l$回目の資産購入時の価格である．また，TCはすべてのアルゴリズムエージェントの資産購入を合計して算出している．アルゴリズムエージェントが存在しない場合，取引価格の平均はファンダメンタル価格とほぼ同じになることが明らかになっている\cite{MMKIKY14}．TCが小さいということは，アルゴリズムエージェントが平均的に低価格で資産を購入できているため，パフォーマンス（運用成績）もよくなると考えられる．

%\section{結果と考察}
\section{Results}
The results of our experiment showed that for the stable market, the relative AA and OAA performances, as characterized 
by which approach yielded lower $TC$, depended on order frequency: the outcome 
for low order frequency was opposite to that for high frequency. For both the 
crash and surge markets, an OAA yielded lower $TC$ than an AA. For the spoof market, 
we found no significant difference in results between AAs and OAAs.

\begin{table}[t]
	\centering\caption{Number of orders, $TC$, and average market prices in stable market\label{antei}}
	\small
	\begin{tabular}{c|cc|cc}
	Order frequency & \multicolumn{2}{c|}{Low} & \multicolumn{2}{c}{High} \\
	& AA & OAA & AA & OAA\\ \hline
	\mbox{\#} of orders	& 284.00  & 284.15 & 867.00 & 866.88 \\
	$TC$		& 23.39 & 19.56 & 55.03 & 65.67 \\
	Average price  & 10,012.24 & 10,017.43 & 10,035.94 & 10,048.91 \\
	\end{tabular}
\end{table}

\begin{figure}[t]
	\includegraphics[width=\linewidth]{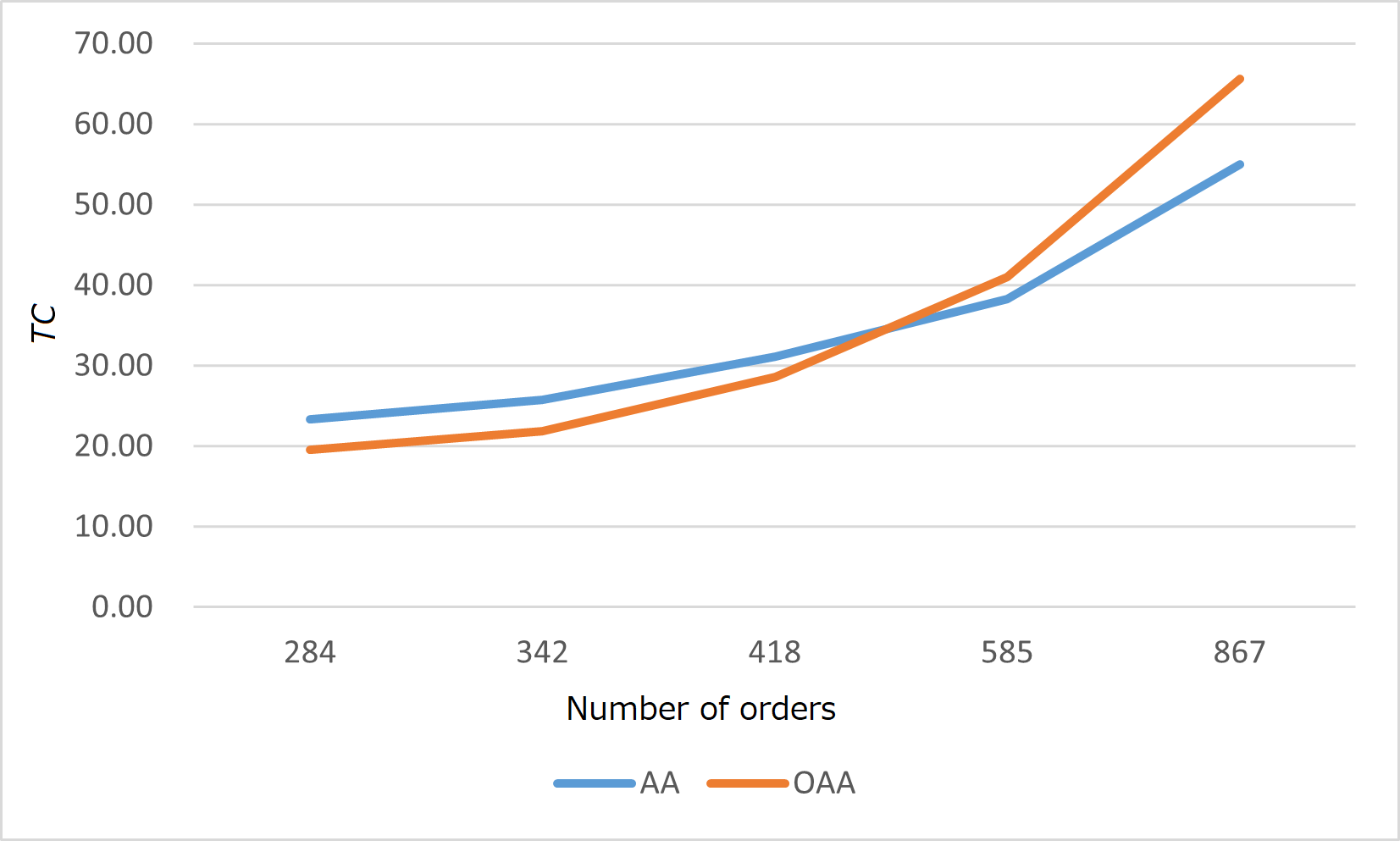}
	\caption{Changes in $TC$ when varying the number of orders placed by algorithm agents\label{TC}}
\end{figure}

For the stable market with algorithm agents executing orders
at low frequency, an OAA yielded lower $TC$
than an AA (Table~\ref{antei}); however, OAAs see a deterioration in performance with
increasing order frequency, producing higher $TC$ than the AAs
for high order frequency (Figure~\ref{TC}).
The reasons for this performance reversal are as follows.
As noted in Section 1, OBI is positively correlated with future
returns; because the OAAs place orders with more precise timing
than the AAs, they should have higher-performance trades.
However, because OAAs take OBI into account, they place
market buy orders immediately before any expected price 
increases; consequently, orders placed 
by the algorithm itself
have the effect of accelerating price increases,
thereby raising the price of the algorithm's
subsequent (buy) orders.
Thus, the ramifications of OAAs include both $TC$-lowering and $TC$-raising effects.
Looking at average market prices (Table~\ref{antei}), we see
that the performance gap between AAs and OAAs widens with increasing
order frequency; thus, at higher order frequencies,
an OAA, despite its ability to place buy orders with precise
timing, suffers from the negative consequences of price increases
induced by its own previous orders, eventually underperforming the AA.\par

\begin{figure}[t]
	\includegraphics[width=\linewidth]{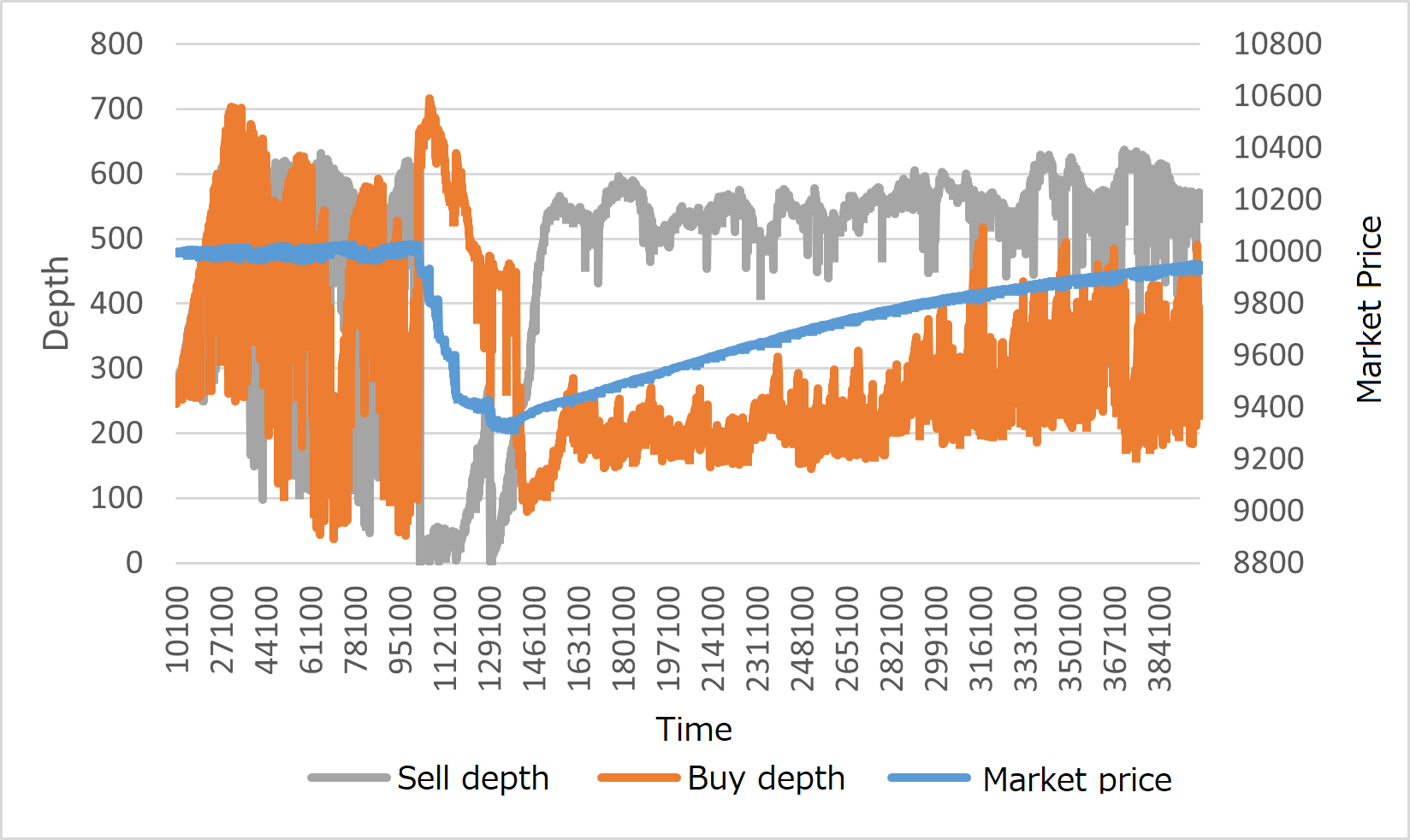}
	\caption{Market price and depth in crash market with OAA participation\label{kyuurakudepth}}
\end{figure}

For both the crash and surge markets,
an OAA yielded lower $TC$ than an AA (Tables~\ref{kyuuraku} and~\ref{kyuutou}). This may be explained by considering
market prices and depths.
In our experiments, we determined buy (sell) depths
by averaging over all intervals between times $t=0$ and $t=t_e$
the total number of orders at prices within a range extending 50
ticks below (above) the best bid (ask).
In the crash market, buy depth exceeds sell depth
over a time interval (designated interval A) that begins when prices start to fall
and ends when prices reach their nadir, just before beginning
to rebound.
Throughout interval A, OAAs consistently place orders to
purchase assets, whereas in later intervals, because sell depth exceeds buy depth,
the algorithm purchases almost no assets (Figure~\ref{kyuurakudepth}).
In contrast, AAs periodically purchase assets throughout all time intervals.
Comparing average market prices over the intervals in which the two
algorithms purchase assets, we find that
the average market price over interval A (during OAA 
participation) is 9,547.15, whereas the average market price over
all intervals (throughout AA participation) is 9,684.63.
Thus, the OAAs succeed in purchasing assets at lower prices than the AAs,
yielding lower $TC$.\par

\begin{figure}[t]
	\includegraphics[width=\linewidth]{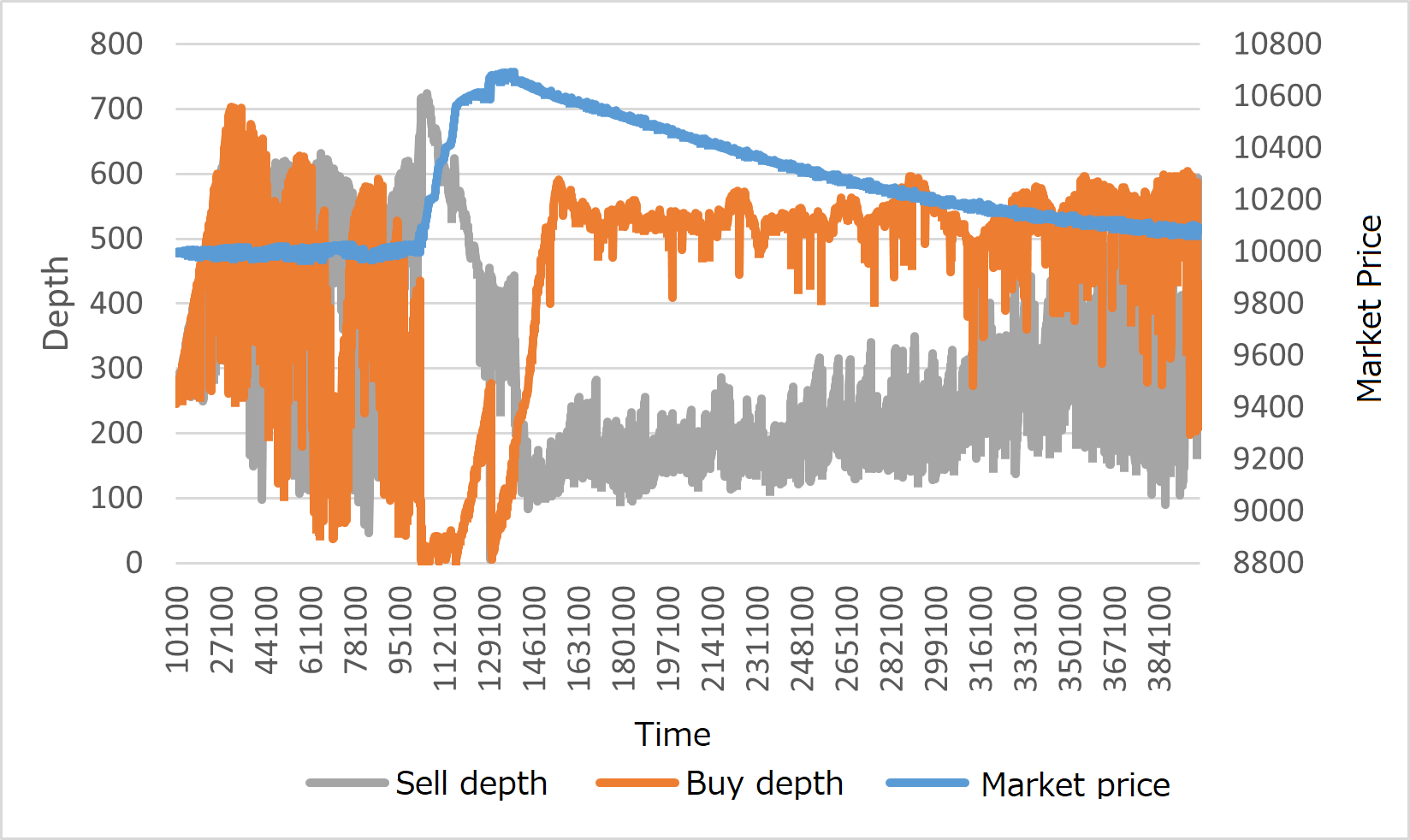}
	\caption{Market price and depth in surge market with OAA participation\label{kyuutoudepth}}
\end{figure}

In the surge market, buy depth exceeds sell depth
primarily over an interval (designated interval B) following the time at which prices
cease to increase and begin falling.
Throughout interval B, the OAAs place orders to purchase assets; however,
before interval B (specifically, from the time prices begin rising
until just before the start of interval B) the OAAs purchase almost no assets,
as sell depth largely exceeds buy depth during this interval 
(Figure~\ref{kyuutoudepth}).
In contrast, as in the case of the crash market,
the AAs continue to purchase assets periodically throughout
all time intervals.
Comparing average market prices over the intervals in which the two
algorithms purchase assets, we find that
the average market price over interval B (during OAA 
participation) is 10,306.90, whereas the average market price over
all intervals (throughout AA participation) is 10,331.95.
Thus, as in the crashing-market case,
the OAAs again succeed in purchasing assets at lower prices than the AAs,
which again yields lower $TC$.\par

For the spoof market,
the $TC$-reducing performance of the OAAs is worse than
that observed in ordinary stable markets
(compare the $TC$ results for the OAAs at low order frequency
in Table~\ref{antei} to the $TC$ results for the OAAs in Table~\ref{misegyoku}).
On the other hand, comparing $TC$ outcomes for the OAAs and AAs
at similar order frequencies in the spoof market
reveals no significant difference in performance (Table~\ref{misegyoku}).
To explain this finding, we note that our spoof market was implemented by 
introducing spoofing manipulations into our ordinary stable
market; thus, during intervals in which no spoofing manipulations
occur, the OAAs obtain lower $TC$ than AAs, as we found in our
stable-market simulation for low order frequency (Table~\ref{antei}).
However, during intervals in which spoofing manipulations
do occur, the OAAs place market buy orders at higher frequency than the AAs,
and thus suffer from the same difficulty observed in our
stable-market simulation for high order frequency, namely,
higher average buy prices with higher $TC$ resulting from price increases
induced by the algorithm's own orders.
Thus, the ramifications of the OAAs include both $TC$-lowering and $TC$-raising
effects
during intervals in which spoofing manipulations are present or
absent, resulting ultimately in little difference between the
performances of the AAs and OAAs.\par

Our results indicate that, for stable markets, the performance of the OAAs worsens
as the frequency of orders placed by algorithm agents increases. As 
noted above, we attribute this to the acceleration of price increases
resulting from the algorithm's own orders.
We suspect that this phenomenon of deteriorating OAA performance
with increasing order frequency may also occur under other
types of market conditions (crash, surge, or spoof markets),
as we observed in this study for stable markets.
Thus, investigating variations in trading costs
with increasing order frequency under crash-, surge-, and spoof-market 
conditions is a key topic for future work.\par

The findings discussed above indicate that
OBI-aware strategies should be incorporated into
execution algorithms.
Indeed, our results demonstrate that, compared to AAs,
OAAs reduce $TC$ under stable-, crash-, and surge-market conditions, 
but do not significantly degrade
performance under spoof-market conditions.
Nevertheless, as the frequency of orders placed by execution
algorithms increases, OBI-aware strategies
suffer from the consequences of market price
increases induced by their own orders, potentially
raising their own subsequent order prices, and thus
caution is necessary when deploying these algorithms in practice.
Conversely, when order frequencies are too low,
the conditions required to trigger orders in OBI-aware strategies
may remain unsatisfied, and thus the length of time required 
to place a given volume of orders may grow unacceptably 
long.\footnote{In our experiment, we used different order 
frequencies for the AAs and OAAs to equalize
the numbers of order submissions. In stable markets, using the same 
order frequencies for the two algorithms has the consequence 
that the time interval required to place a given number
of orders is approximately twice as long for the OAAs than for the AAs,
because the OAAs place orders only during intervals in which
buy depth exceeds sell depth.}

\begin{table}[t]
	\centering\caption{Numbers of orders and MI in crash market\label{kyuuraku}}
	\begin{tabular}{c|cc}
	& AA & OAA\\ \hline
	\# of orders & 83.00  & 82.65 \\
	MI		& -321.65  & -431.80  \\
	\end{tabular}
\end{table}
\begin{table}[t]
	\centering\caption{Number of orders and MI in surge market\label{kyuutou}}
	\begin{tabular}{c|cc}
	& AA & OAA\\ \hline
	\# of orders	& 511.00 & 511.23 \\
	MI		& 337.54 & 313.02 \\
	\end{tabular}
\end{table}
\begin{table}[t]
	\centering\caption{Numbers of orders and MI in spoofing market\label{misegyoku}}
	\begin{tabular}{c|cc}
	& AA & OAA\\ \hline
	\# of orders	& 443.00 & 443.12 \\
	MI		& 31.30 & 32.60 \\
	\end{tabular}
\end{table}


\section{Conclusion}
In this study, to investigate the influence of market conditions on OBI-aware
execution algorithms, we simulated artificial markets reflecting a variety of
distinct market environments and compared the trading costs obtained
using AAs and OAAs.
For stable markets in which algorithm agents place orders at low frequency, our results
demonstrated lower trading cost for the OAAs than for the AAs.
On the other hand, for stable markets with high order frequency, the OAAs suffer
from the consequences of market price increases induced by their own orders, resulting in 
increased trading cost.
In both crashing and surging markets, the OAAs obtain lower trading costs than the AAs.
In a spoofing market, we found no significant difference between trading costs
obtained using AAs and using OAAs.
However, one point requiring caution is that the phenomenon we
observed for stable markets, that trading cost for OAAs increases with increasing
order frequency, may also arise for crash, surge, or spoof markets.\par
The results of this study indicate that the introduction
of OBI-aware strategies may enable execution algorithms
to trade more effectively.
However, our findings also indicate one potential challenge:
identifying optimal order frequencies for these algorithms
is complicated, with consequences ranging from
increased market prices induced by the algorithm's own orders
to unacceptably long time intervals required to complete
a given volume of trades.\par
Our focus in this study was restricted to the narrow question of
whether OBI-aware strategies could enhance the performance
of execution algorithms. For this reason, we chose a simplified
simulation model satisfying the minimal set of constraints required
to exhibit clear distinctions in algorithm performance.
In particular, our model excluded scheduling considerations, such
as the question of how best to conclude the execution of a full trade, and
was thus unable to assess the optimality of execution schedules
as a whole. In future work, we will extend our study to incorporate
more general models of execution algorithms.
Similarly, our study of crash, surge, and spoof markets employed
only the minimal models needed to incorporate the market 
conditions in question. An additional topic for future work is
to adopt more realistic models of market conditions based on data
from actual markets, including not only price crashes and surges
but also price fluctuations and order volumes in cases involving spoofing.
This study identified certain conditions under which OAAs had
higher trading costs than AAs; it is interesting to ask under what
other sorts of conditions this phenomenon might occur,
as well as how the performance of OAAs is affected by the behaviour of other market
participants. Finally, our consideration of spoofing was restricted
to the case of spoofing manipulations of stable markets; it would be
interesting to investigate the consequences of spoofing in other types
of market environment.

\section*{Funding Statement}
This work was supported by JSPS KAKENHI [grant number 23K04276].
The authors thank FORTE Science Communications (https://www.forte-science.co.jp/) for English language editing.
\section*{Declarations}
It should be noted that the opinions contained herein are solely those
of the authors and do not necessarily reflect those of SPARX Asset
Management Co., Ltd.

\bibliography{bibFIN23-Ie.bib}
\bibliographystyle{plain}

% \appendix
%\begin{biography}
%%
%\profile{n}{遠藤 修斗}{2023年4月より工学院大学大学院工学研究科システムデザイン専攻に在籍．マルチエージェントシミュレーションの金融分野応用研究に興味を持つ．}{endo}
%\profile{m}{水田 孝信}{2000年気象大学校卒業．2002年東京大学大学院理学系研究科修了．2004年同研究科博士課程中退．同年4月スパークス・アセット・マネジメント株式会社入社．2010年5月よりファンドマネージャー．2014年9月東京大学大学院工学系研究科システム創成学専攻博士課程修了．博士（工学）．日本証券アナリスト協会検定会員．進化経済学会，JAFEE(日本金融・証券計量・工学学会)各会員．}{mizuta}
%\profile{m}{八木 勲}{	1995年大阪大学基礎工学部情報工学科卒業．1997 年奈良先端科学技術大学院大学情報科学研究科博士前期課程修了．日立造船（株）などを経て，2006 年奈良先端科学技術大学院大学情報科学研究科博士後期課程修了．博士（工学）．東京工業大学大学院総合理工学研究科特別研究員を経て，2011 年より神奈川工科大学情報学部准教授．2021年より工学院大学情報学部システム数理学科教授．金融，経済，および教育分野に関する社会シミュレーションに興味を持つ．電子情報通信学会，情報処理学会，進化経済学会各会員．}{yagi}
%% 
%\end{biography}
\end{document}